# Unified and Efficient Analysis of Machining Chatter and Surface Location Error

Woraphrut Kornmaneesang, Tsu-Chin Tsao, Niloufar Esfandi, and Shyh-Leh Chen

*Abstract*— **Although machining chatter can be suppressed by the choice of stable cutting parameters through means of stability lobe diagram (SLD), surface roughness still remains due to the forced vibration, which limits surface quality, especially in the surface finish. Better cutting parameters can be achieved considering surface location error (SLE) together with SLD. This paper proposes an innovative modeling framework of the machining dynamic system that enables efficient computation of the chatter stability and SLE. The framework mainly embodies two techniques, namely semi-discretization method (SDM) and lifting method. The machining dynamics system is mathematically expressed as an angle-varying delay differential equation (DDE). The SDM approximates the angle-varying and delayed terms to ordinary terms using zero-phase interpolations and governs the discrete angle-varying dynamics system. Then, the system is merged over the tooth passing angle using the lifted approach to establish an explicit dynamic system in the compact state-space form. Based on the compact state-space model, the chatter stability and SLE prediction are easily and efficiently conducted. Simulation results show the improved efficiency of the proposed method over other well-known methods.**

## I. INTRODUCTION

MACHINING is one of the most prevalent manufacturing processes in various industrial domains, including aerospace, automotive, semiconductor, electronic, and medical [1]. The machining forces create mechanical vibrations, which can be classified into stable forced vibrations and unstable self-excited oscillations, also known as chatter. Both play a key role in limiting production efficiency, surface quality, and tool life. Thus, it is important to avoid or suppress the chatter problems, while establishing acceptable levels of surface quality.

The onset of chatter signifies unstable cutting due to the regenerative feedback effects of the cutting tool engaging in a part of surface that has been cut in the previous cutter path [2]. Stability lobe diagram (SLD), which indicates the stability regions with respect to the cutting load and speed, provides important information for selecting cutting parameters to avoid chatter [3]. Analysis of machining stability was firstly studied by Tobias and Fishwick [4] and Tlusty [5]. Tobias introduced a method of generating stability lobes, which present critical axial depths of cut in relation to spindle speeds, to isolate stable and unstable cutting regions. As this classical method is established by phase reasoning in the frequency-domain and is limited to single-point cutting tool operations, such as turning and boring, many attempts have been made based on rigorous mathematical models and also to extended to more general cutting operations like milling. They can be classified into frequency-domain Fourier series truncation, time-domain finite difference, and time-domain discretization methods.

Altintas and Budak proposed an analytical frequency-domain-based method [6]. Time-periodic terms are reduced to zeroth-order constant terms using the Fourier series expansions; thus, this method was called zeroth-order approximation (ZOA) or single-frequency method. Despite a rapid estimation of the stability, there exists inaccuracy in highly interrupted cutting processes with small radial immersions. To overcome this limitation, Merdol and Altintas introduced a multi-frequency method that includes higher harmonics of the time-periodic terms, resulting in better estimation accuracy [7]. Meanwhile, Landers and Ulsoy [8] and Munoa and Yang [9] developed semi-analytical approaches from the ZOA method, allowing the nonlinear cutting force and the interrupted cutting to be included. However, the inclusion of higher harmonics significantly increases the computational complexity of the chatter frequency scanning process, introducing a trade-off between accuracy and computation efficiency. More importantly, these frequency-domain methods may be limited by the uncertain frequency spill-over effect on the Fourier series truncation interacting with the time delay phase shift in more complex cutting conditions, such as those involved highly-deformable workpieces and variable-pitch cutting tools.

Time-domain mathematical models, which include process damping, structural and cutting force nonlinearities, as well as complex tool geometries, were modeled and analyzed as delayed differential equations (DDEs) with periodic coefficients [10, 11]. Stability analysis for the periodic DDEs in the time domain has been solved by the finite difference approximation, including Euler, Runge-Kutta, and Tustin numerical methods [12-16]. However, the finite difference approximations of the entire DDEs require rather fine time steps to render numerical accuracy, and as such incur a heavy computation load.

Woraphrut Kornmaneesang is with the department of Mechatronic Engineering, National Taiwan Normal University, Taipei 106, Taiwan (email: woraphrut.korn@ntnu.edu.tw).

Tsu-Chin Tsao is with the department of Mechanical and Aerospace Engineering, UCLA, Los Angeles, CA 90095 (email: ttsao@ucla.edu).

Shyh-Leh Chen is with the department of Mechanical Engineering and AIM-HI, National Chung Cheng University, Chiayi 621, Taiwan (email: imeslc@ccu.edu.tw).

Niloufar Esfandi was with the department of Mechanical and Aerospace Engineering, UCLA, Los Angeles, CA 90095 (email: niloufaresfandi@ucla.edu).

The time-domain discretization methods discretize the DDEs into finite intervals and apply the Floquet theory for the stability analysis, recognized as semi-analytical approaches, significantly improving computation efficiency over the finite difference numerical methods. Moreover, the analysis through the DDEs enables the time-domain methods to consider more complex cutting conditions than the frequency domain methods. Insperger and Stépán presented the semi-discretization method (SDM) which discretizes the DDEs by approximating the time-varying terms with piecewise constants and the delayed terms with polynomial interpolations, while the rest terms are unchanged [17]. Variations to the SDM evolve around the interpolation to render improved numerical accuracy. For example, linear interpolation to both the delay and time-varying terms (FDM) [18], higher-order polynomial interpolations – 1st SDM [19], 2nd SDM [20, 21], 2nd FDM [22], and 3rd updated FDM (UFDM) [23], least squares interpolation [24, 25], Newton interpolation [23, 26]. Another improvement exploits the nature of interrupted cutting by treating the air cutting phase as one segment with free vibration solution and the immersed cutting phase, during which the delayed and varying terms reside, as either uniformly discretized segments solved by the numerical integration methods (NIMs) [27-29] or non-uniformly discretized segments solved by the temporal finite element analysis (TFEA) [30] and the Chebyshev collocation method (CCM) [31, 32]. This improves the efficiency of the stability analysis for low-immersion cutting.

In analyzing forced vibration in stable cutting, several researches reconstructed the machined surface from the tool/workpiece relative motion [33-36]. However, their algorithms are too complex and time-consuming. Schmitz and Ziegert introduced surface location error (SLE), calculated directly from the forced vibrations, instead of surface roughness or surface shape, to facilitate the surface quality assessment [37]. SLE is defined as the maximal distance between the desired surface and the machined surface, which can be estimated by the steady-state tool/workpiece vibratory motion with the simplified tool geometry. Schmitz and Mann calculated the SLE through the frequency domain [38]. Meanwhile, Insperger et al. adopted the harmonic balance method incorporated with the Fourier series expansion to predict the SLE [39, 40]. Nevertheless, these methods are limited to only predicting SLE. Later, several research groups have extended the time-domain discretization methods to predict not only the chatter stability, but also the SLE. Mann et al. presented simultaneous prediction of chatter stability and SLE using TFEA [41, 42]. Ding et al. extended his previous work, i.e., FDM, with the precise time integration (PTI) method for higher computational efficiency [43]. Li et al. conducted the stability and SLE predictions based on SDM, including the effects of mode coupling and process damping [20].

Compared to the first two approaches, the time-domain discretization methods offer improved computational efficiency and, moreover, enable simultaneous estimation of chatter stability and SLE. However, the stability and SLE predictions require calculating the monodromy (or Floquet transition) matrix eigenvalues and inverse, respectively, which are the predominant portion of the numerical computation. In fact, the computational time grows proportionally to the cube of the matrix dimension $\mathcal{O}(\eta^3)$, where $\eta$ is the monodromy matrix dimension [44]. Modern machining with complex cutting conditions, such as multiple participation modes of vibration, coupled modes, and variable pitch cutters, demand the greater number of states that render a larger monodromy matrix dimension. Consequently, such complex cutting imposes a heavy computational load in the stability and dynamic analysis.

This paper presents a novel approach to minimize the monodromy matrix dimension, accelerating the numerical computations in the time-domain discretization methods. Instead of directly discretizing the closed-loop DDEs of the machining dynamics, the cutting force and mechanical models are separately discretized by zero-phase SDM and continuous-to-discrete conversion methods from the control theory, respectively. Both the discretized models are then lifted over the tooth-passing period and combined through the feedback structure to form a minimal state-space representation of the closed-loop machining dynamics. This formulation achieves a reduced monodromy matrix dimension, significantly speeding up the stability analysis and SLE prediction. As such, the following contributions are highlighted:

(i) The proposed formulation achieves a minimal monodromy matrix dimension of $r(2n+m)$ considering the input-output dynamics, in contrast to $r(2n)(m+1)$ in the existing discretization methods, where $r$, $n$, and $m$ denote the numbers of motion axes, participation modes, and discretization intervals, respectively. This dimensional reduction noticeably enhances computational efficiency of the chatter stability analysis, particularly in scenarios involving complex cutting conditions.

(ii) The zero-phase SDM is presented to eliminate the phase distortion, which is inherently introduced in the existing interpolation methods. As such, fewer discretization steps than other SDM methods are required for the stability analysis and SLE prediction.

(iii) The proposed framework enables the steady-state vibration calculation using a minimal-state feedback closed-loop dynamic system, where no monodromy matrix calculation is involved. This considerably improves computational efficiency of the SLE prediction.

Note that the proposed treatment can be integrated with other time-domain discretization methods, such as SDMs/FDMs, NIM, TFEA, and CCM, to further enhance their computational accuracy and efficiency.

The remainder of this paper is organized as follows. The milling modeling is introduced in Section II, while the proposed method for the stability analysis and SLE prediction is presented in Section III. Then, computations and analyses are shown in Section IV. The conclusions are finally outlined in Section V.

## II. MILLING DYNAMIC MODEL

To analyze the stability and predict the SLE, the angle-periodic DDE of the machining dynamics is needed. It can be broken into two submodels, i.e., cutting force model and mechanical model. The angle domain, instead of the time domain, is formulated because the cutting force geometry and dynamics are directly related to the cutter angles. Hence, in this section, the standard 2-dimensional cutting force model of milling processes is introduced, and then a 2-dimensional $n$-mode mechanical model is addressed.

### A. Cutting Force Model

The mathematical multi-tooth milling cutting force has been established in [45] and broadly used in the literature. In the following, the mathematical model with two orthogonal degrees of freedom (DOFs) in the $X$ and $Y$ directions shown in Fig. 1 is rewritten in compact general vector-matrix forms to facilitate the subsequent dynamic analysis.

The cutter is assumed to have $N$ number of identically spaced teeth with a zero-helix angle. The cutting force $\boldsymbol{f}_j$ exerting to the $j$-th cutter teeth in the two coordinate systems, namely fixed Cartesian ($xy$) and tangential-normal ($tn$), are defined below:

$$\boldsymbol{f}_j^{xy} = \begin{bmatrix} f_j^x(\phi_j) \\ f_j^y(\phi_j) \end{bmatrix}, \quad \boldsymbol{f}_j^{tn} = \begin{bmatrix} f_j^t(\phi_j) \\ f_j^n(\phi_j) \end{bmatrix} \tag{1}$$

with

$$\boldsymbol{f}_j^{xy} = \boldsymbol{R}(\phi_j)\boldsymbol{f}_j^{tn}; \quad \boldsymbol{R}(\phi_j) = \begin{bmatrix} -\cos\phi_j & -\sin\phi_j \\ \sin\phi_j & -\cos\phi_j \end{bmatrix}$$

$$\phi_j(\theta) = \theta + \frac{2\pi}{N}(j-1); \quad j = 1,2,\dots,N$$

where the independent variable $\theta = \frac{2\pi\Omega}{60}t$ is the spindle angle; $\Omega$ is the spindle speed. Let us define $\Delta z = [\Delta x \quad \Delta y]^T$ as the relative displacement between the cutting tool and the workpiece in the Cartesian coordinate frame.

The tangential and normal cutting forces at the $j$-th tooth are related to the axial depth of cut $a_p$ and the instantaneous chip thickness $h(\phi)$ by the proportional cutting coefficients $\boldsymbol{k}_c = [k_c^t \quad k_c^n]^T$ as well as $\boldsymbol{k}_e = [k_e^t \quad k_e^n]^T$, modulated by the engagement factor $g(\phi_j)$:

$$\boldsymbol{f}_j^{tn}(\phi_j) = a_p g(\phi_j)\left[\boldsymbol{k}_c h(\phi_j) + \boldsymbol{k}_e\right] \tag{2}$$

with

$$h(\phi_j) = -\begin{bmatrix} 0 & 1 \end{bmatrix} \boldsymbol{R}^{-1}(\phi_j)\left[\boldsymbol{s}_t + \boldsymbol{\Delta z}(\theta) - \boldsymbol{\Delta z}(\theta-\Theta)\right] \tag{3}$$

The instantaneous chip thickness $h$ consists of the static part, i.e., the feed per tooth vector $\boldsymbol{s}_t = [s_x \quad s_y]^T = \frac{30\Theta}{\pi\Omega}\boldsymbol{v}$ in the machine tool's Cartesian coordinate frame, and the dynamic part, which includes the regenerative effects of the delayed feedback by one tooth passing angle $\Theta = \frac{2\pi}{N}$, where $\boldsymbol{v} = [v_x \quad v_y]^T$ is the feed motion. In the existing literature, feed per tooth vector $\boldsymbol{s}_t$ has been considered a scalar variable, i.e., the feed direction aligns with one of the machine tool's feed axes, in the existing literature. Therefore, $\boldsymbol{s}_t = [s_x \quad 0]^T$ for the $x$-axis movement is illustrated in Fig. 1. The vector notation here is more general and useful when asymmetric dynamics exhibit in the machine tool's feed drive and structural dynamics. The force in the cutter axial direction can also be readily included but is omitted herein to be consistent with most existing literature. The function $g(\phi_j)$ is a switching function that determines whether the $j$th tooth is in or out of cut:

$$g(\phi_j) = \begin{cases} 1; & \phi_{st} \le \phi_j \le \phi_{ex} \\ 0; & \text{otherwise} \end{cases} \tag{4}$$

where $\phi_{st}$ and $\phi_{ex}$ are the start and exit immersion angles of the cutter, respectively. They are given as:

$$\begin{aligned} \text{Up-milling} & \begin{cases} \phi_{st} = 0 \\ \phi_{ex} = \cos^{-1}(1-2a_r/D) \end{cases} \\ \text{Down-milling} & \begin{cases} \phi_{st} = \cos^{-1}(2a_r/D-1) \\ \phi_{ex} = \pi \end{cases} \end{aligned} \tag{5}$$

where $a_r$ is the radial depth of cut and $D$ is the cutter diameter.

Summing the forces contributed from all the teeth and transform to the Cartesian frame, one can get:

$$\begin{aligned} \boldsymbol{f}^{xy}(\theta) &= \sum_{j=1}^{N} \boldsymbol{f}_j^{xy}(\theta) \\ &= a_p \left\{ \underbrace{\begin{bmatrix} r_1(\theta) \\ r_2(\theta) \end{bmatrix}}_{r(\theta)} - \underbrace{\begin{bmatrix} s_{11}(\theta) & s_{12}(\theta) \\ s_{21}(\theta) & s_{22}(\theta) \end{bmatrix}}_{S(\theta)} \left[\boldsymbol{s}_t + \boldsymbol{\Delta z}(\theta) - \boldsymbol{\Delta z}(\theta-\Theta)\right] \right\} \end{aligned} \tag{6}$$

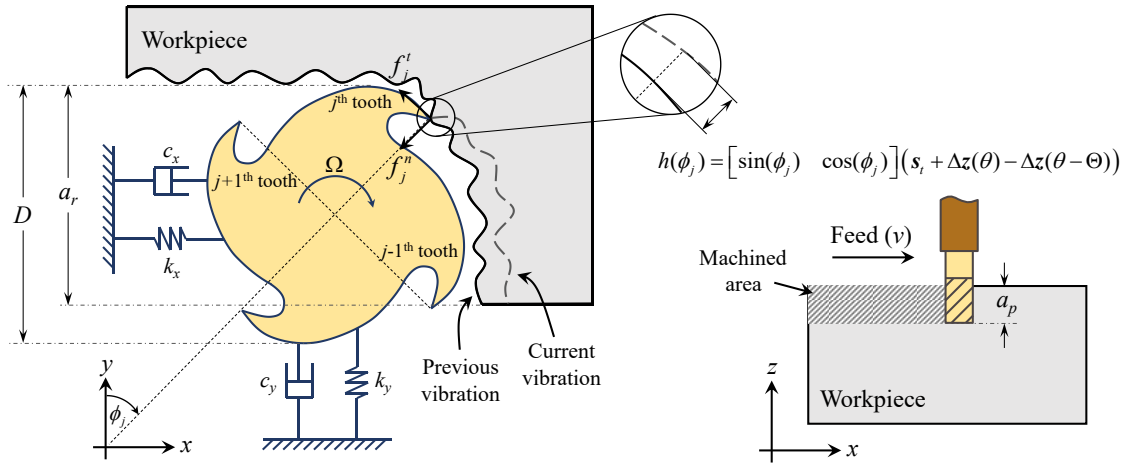

Fig. 1. A schematic diagram of a single-mode 2-dimensional milling dynamic system

where

$$r_1(\theta) = -\sum_{j=1}^{N} g(\phi_j)\left( k_e^t \cos\phi_j + k_e^n \sin\phi_j \right)$$

$$r_2(\theta) = \sum_{j=1}^{N} g(\phi_j)\left( k_e^t \sin\phi_j - k_e^n \cos\phi_j \right)$$

$$s_{11}(\theta) = \frac{1}{2}\sum_{j=1}^{N} g(\phi_j)\left[ k_c^t \sin 2\phi_j + k_c^n (1-\cos 2\phi_j) \right]$$

$$s_{12}(\theta) = \frac{1}{2}\sum_{j=1}^{N} g(\phi_j)\left[ k_c^t (1+\cos 2\phi_j) + k_c^n \sin 2\phi_j \right]$$

$$s_{21}(\theta) = \frac{1}{2}\sum_{j=1}^{N} g(\phi_j)\left[ -k_c^t (1-\cos 2\phi_j) + k_c^n \sin 2\phi_j \right]$$

$$s_{22}(\theta) = \frac{1}{2}\sum_{j=1}^{N} g(\phi_j)\left[ -k_c^t \sin 2\phi_j + k_c^n (1+\cos 2\phi_j) \right]$$

The terms $r(\theta)$ and $S(\theta)$ in (6) are often called periodic dynamic force coefficients, which have a period of the tooth passing angle $\Theta$ and are contributed from the cutting coefficients and the immersion ratio $a_e/D$ [46]. Without loss of generality, the term $f^{xy}$ is simplified to $f$ for the following derivation.

### B. Mechanical Dynamic Model

Mechanical structural dynamics obtained from finite element modeling, field test modal analyses or combined models can be represented by a state space presentation of an $r$-dimensional axis with $n$-mode (including both workpiece and cutting tool sides) 2nd-order mechanical dynamics on each dimensional axis as:

$$\dot{q}(t) = Aq(t) + Bf(t)$$
$$\Delta z(t) = Cq(t) \tag{7}$$

where $q \in \Re^{2rn}$ is the state vector and $A \in \Re^{2rn \times 2rn}$, $B \in \Re^{2rn \times r}$, and $C \in \Re^{r \times 2rn}$ are the state space matrices. Note that Fig. 1 illustrates the 2-dimensional single-mode ($r = 2$, $n = 1$) mechanical model.

Here, the cutting force model (6) and the mechanical model (7) are expressed in different domains. They must be in the same domain for further analysis. The angle domain presents the models with respect to the cutter angular position, carrying a more intuitive meaning than the time domain. By the chain rule, one can get:

$$\dot{q}(t) = \frac{dq(t)}{dt} = \frac{dq}{d\theta}\frac{d\theta}{dt} = q'(\theta)\omega \tag{8}$$

where $\omega = \frac{2\pi}{60}\Omega$. Hence, the mechanical model in the angle domain is obtained as:

$$q'(\theta) = A_\omega q(\theta) + B_\omega f(\theta)$$
$$\Delta z(\theta) = C_\omega q(\theta) \tag{9}$$

where $A_\omega = A/\omega$, $B_\omega = B/\omega$, and $C_\omega = C$.

Combining the cutting process model (6) with the generalized mechanical system (9) formulates a closed-loop machining system. Its block diagram is shown in Fig. 2(a).

### III. Stability and Surface Location Error Analysis

The system, as depicted by Fig. 2(a), may be used to simulate the stability and forced vibration (surface profile

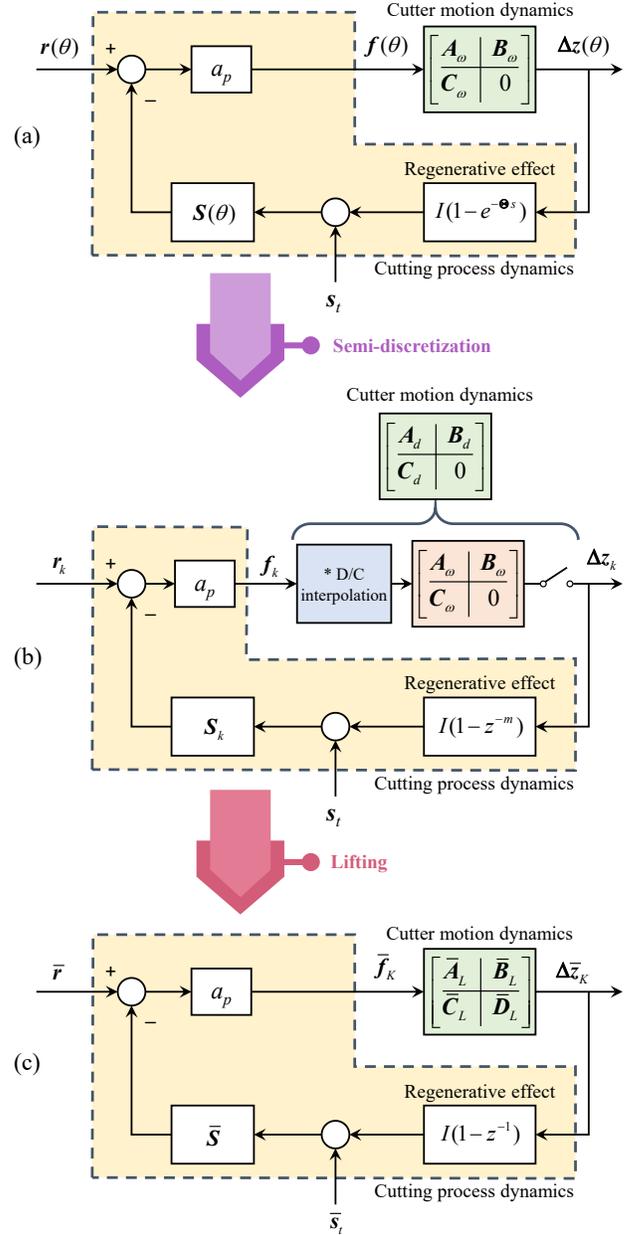

Fig. 2 Block diagrams presenting overall interactions between the mechanical system and the cutting process: (a) continuous-angle system, (b) semi-discretized system, and (c) lifted semi-discretized system

error) subject to a given feed per tooth $s_t$. For parametric analysis, the angle-periodic and DDE prohibits direct linear shift-invariant system analysis. In this section, the Floquet theorem-based analysis used in the existing methods is revisited for the purpose of comparison with the proposed method. Then, the zero-phase semi-discretization and the lifting method are presented. This is followed by a stability analysis and the SLE prediction of the milling system.

### A. Review of Existing Analysis Methods

All the existing time-domain discretization methods govern the closed-loop milling dynamics by combining (6) and (9) into a general DDE form:

$$\boldsymbol{q}'(\theta) = (\boldsymbol{A}_p - \boldsymbol{B}_p(\theta))\boldsymbol{q}(\theta) + \boldsymbol{B}_p(\theta)\boldsymbol{q}(\theta - \Theta) + \boldsymbol{w}(\theta) \quad (10)$$

where $\boldsymbol{A}_p = \boldsymbol{A}_\omega$; $\boldsymbol{B}_p(\theta) = a_p \boldsymbol{B}_\omega \boldsymbol{S}(\theta)\boldsymbol{C}_\omega$; $\boldsymbol{w}(\theta) = a_p \boldsymbol{B}_\omega[\boldsymbol{r}(\theta) - \boldsymbol{S}(\theta)\boldsymbol{s}_t]$. Discretizing the DDE (10) and apply the Floquet theory formulate a common discrete state equation:

$$\boldsymbol{\xi}_{a,K} = \boldsymbol{\Phi}_a \boldsymbol{\xi}_{a,K-1} + \boldsymbol{\sigma}_a \quad (11)$$

where $\boldsymbol{\xi}_{a,K}$ and $\boldsymbol{\xi}_{a,K-1}$ are the augmented discretized states, including vibratory displacement and velocity, within the current and previous tooth passing angle, respectively. The state definition varies according to the discretization technique. $\boldsymbol{\Phi}_a$ and $\boldsymbol{\sigma}_a$ are the monodromy matrix and the forced vibration term contributed from the cutting conditions. The stability is determined by solving the spectral radius of the monodromy matrix $\boldsymbol{\Phi}_a$, while the SLE is calculated using the steady-state vector $\boldsymbol{\xi}_a^{ss} = (\mathrm{I} - \boldsymbol{\Phi}_a)^{-1}\boldsymbol{\sigma}_a$.

However, the computational cost of the monodromy matrix eigenvalue and inverse grows proportionally to the cube of the matrix dimension $\mathcal{O}(\eta^3)$ [44]. In most existing time-domain discretization methods, the monodromy matrix dimension is $r(2n)(m+1)$. Although the SDMs/FDMs can reduce the matrix dimension to $rn(m+2)$ by removing delayed velocity states [17, 18], their computational load remains heavy due to the iterative matrix multiplications in the monodromy matrix computation.

### B. Zero-phase Semi-discretization

The existing methods directly implement the discretization techniques on the closed-loop dynamics (10), and perform the stability and vibration analysis using the means of the Floquet theory. Meanwhile, the proposed analysis method applies the zero-phase semi-discretization on the cutting force dynamic and the mechanical system, separately, and then, constructs the lifted closed-loop dynamic for the analysis, as presented in the following.

Referring to [17, 46], semi-discretization of a system is to discretize part of the system by approximating the angle-periodic and delay terms with piecewise constant functions and discretized points, respectively, into $m$ intervals ($\Theta = m\Delta\theta$). Thus, the cutting force model (6) is expressed in the discrete-angle form as:

$$\boldsymbol{f}_k = a_p \left[ \boldsymbol{r}_k - \boldsymbol{S}_k \left( \boldsymbol{s}_t + \Delta\boldsymbol{z}_k - \Delta\boldsymbol{z}_{k-m} \right) \right], \quad k = 0,1,\dots \quad (12)$$

where $\boldsymbol{f}_k = \boldsymbol{f}(k\Delta\theta)$ and $\Delta\boldsymbol{z}_k = \Delta\boldsymbol{z}(k\Delta\theta)$. To minimize the estimation inaccuracy due to the phase delay, the periodic constant functions are defined as the average over one sampling interval around the discretized points:

$$\begin{aligned} \boldsymbol{r}_k &= \boldsymbol{r}_{k-m} = \frac{1}{\Delta\theta}\int_{(k-\frac{1}{2})\Delta\theta}^{(k+\frac{1}{2})\Delta\theta} \boldsymbol{r}(\theta)d\theta; \\ \boldsymbol{S}_k &= \boldsymbol{S}_{k-m} = \frac{1}{\Delta\theta}\int_{(k-\frac{1}{2})\Delta\theta}^{(k+\frac{1}{2})\Delta\theta} \boldsymbol{S}(\theta)d\theta; \end{aligned} \quad (13)$$

The semi-discretization on the part of the signals (as opposed to coefficients), from the signal and system's perspective, is equivalent to a discrete to continuous conversion with interpolations by zero-order holding, first-order holding, or a higher-order polynomial holding function in the system's signal flow path, as shown in Fig. 2(b). Such interpolation has inherent phase delay, e.g., one-half of the sampling period for the zero-order hold piecewise constant interpolation. The delay in the feedback system makes the stability determination inaccurate, unless a sufficiently small sampling period is applied. To address this inherent delay in the interpolation, the reconstruction holding function must not introduce phase in the frequency domain. This means that the impulse response must be symmetric with respect to time at zero. We introduce two zero-phase continuous-to-discrete conversions, namely impulse invariance (IMP) and zero-order hold (ZOH) for piecewise constant interpolation, where the continuous domain input is respectively reconstructed by the discretely sampled signal as follows:

$$\text{IMP:} \quad \boldsymbol{f}(\theta) \approx \boldsymbol{f}_k \delta(\theta - k\Delta\theta)\Delta\theta \quad (14)$$

$$\text{ZOH:} \quad \boldsymbol{f}(\theta) \approx \begin{cases} \boldsymbol{f}_k, & k\Delta\theta \le \theta < (k+\tfrac{1}{2})\Delta\theta \\ \boldsymbol{f}_{k+1}, & (k+\tfrac{1}{2})\Delta\theta \le \theta < (k+1)\Delta\theta \end{cases} \quad (15)$$

where $k\Delta\theta \le \theta < (k+1)\Delta\theta$ and $\delta(\cdot)$ is the unit impulse function. The discrete signal is simply the direct sampling from the continuous domain (6), as opposed to the integration of the piecewise constant in (12). The zero-phase holds do not introduce phase shift in the conversion, so that the discrete-angle system characteristics, especially stability, are better preserved.

Applying the sampling at the tool's dynamic displacement and reconstruct the discrete cutting force as continuous input to the structural dynamics as in Fig. 2(b), an exact discrete model for the structural dynamics can be derived as:

$$\begin{aligned} \boldsymbol{p}_{k+1} &= \boldsymbol{A}_d \boldsymbol{p}_k + \boldsymbol{B}_d \boldsymbol{f}_k \\ \Delta\boldsymbol{z}_k &= \boldsymbol{C}_d \boldsymbol{p}_k + \boldsymbol{D}_d \boldsymbol{f}_k \end{aligned} \quad (16)$$

where $(\boldsymbol{A}_d, \boldsymbol{B}_d, \boldsymbol{C}_d, \boldsymbol{D}_d)$ are related to the continuous dynamics (9) in analytical forms:

$$\boldsymbol{p}_k = \boldsymbol{q}_k - \boldsymbol{E}_d \boldsymbol{f}_k \quad (17)$$

and

$$\text{IMP:} \begin{cases} \boldsymbol{A}_d = e^{\boldsymbol{A}_\omega \Delta\theta} & \boldsymbol{B}_d = \boldsymbol{A}_d \boldsymbol{B}_\omega \Delta\theta \\ \boldsymbol{C}_d = \boldsymbol{C}_\omega & \boldsymbol{D}_d = 0 \\ \boldsymbol{E}_d = 0 \end{cases} \quad (18)$$

$$\text{ZOH:} \begin{cases} \boldsymbol{A}_d = e^{\boldsymbol{A}_\omega \Delta\theta} & \boldsymbol{B}_d = e^{\boldsymbol{A}_\omega \frac{\Delta\theta}{2}}\left[\boldsymbol{A}_d - \boldsymbol{I}\right]\boldsymbol{A}_\omega^{-1}\boldsymbol{B}_\omega \\ \boldsymbol{C}_d = \boldsymbol{C}_\omega & \boldsymbol{D}_d = \boldsymbol{C}_\omega \boldsymbol{E}_d \\ \boldsymbol{E}_d = (e^{\boldsymbol{A}_\omega \frac{\Delta\theta}{2}} - \boldsymbol{I})\boldsymbol{A}_\omega^{-1}\boldsymbol{B}_\omega \end{cases} \quad (19)$$

The derivations of (18)-(19) can be found in Appendix A. The discrete structural dynamics (16) and the sampled delayed feedback with periodically varying gains in (12) form a sampled data linear periodic feedback system.

### C. Minimal State-space Realization with Lifting Method

To capture all the states over one period, the lifting method is applied to convert the periodic varying metal cutting system to a linear shift-invariant system [47-49]. First, on the linear shift-invariant structural dynamics (22), the vibration over the tooth passing angle $\Theta$ is:

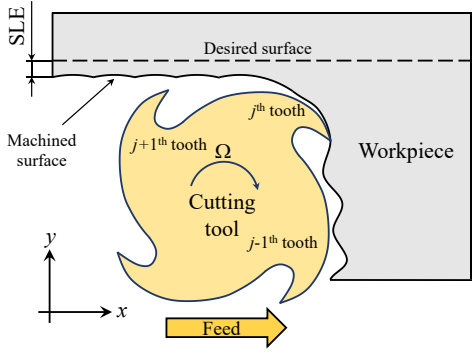

Fig. 3. A schematic diagram of the surface location error

$$\overline{\boldsymbol{p}}_{K+1} = \overline{\boldsymbol{A}}_L\, \overline{\boldsymbol{p}}_K + \overline{\boldsymbol{B}}_L\, \overline{\boldsymbol{f}}_K$$
$$\Delta \overline{\boldsymbol{z}}_K = \overline{\boldsymbol{C}}_L\, \overline{\boldsymbol{p}}_K + \overline{\boldsymbol{D}}_L\, \overline{\boldsymbol{f}}_K \tag{20}$$

where $\overline{\boldsymbol{p}}_K = \boldsymbol{p}_{mk}$; $\overline{\boldsymbol{p}}_{K+1} = \boldsymbol{p}_{m(k+1)}$;

$\overline{\boldsymbol{f}}_K = \left[ \boldsymbol{f}_{mk}^T \quad \boldsymbol{f}_{mk+1}^T \cdots \ \boldsymbol{f}_{mk+m-1}^T \right]^T$; $\Delta \overline{\boldsymbol{z}}_K = \left[ \Delta \boldsymbol{z}_{mk}^T \quad \Delta \boldsymbol{z}_{mk+1}^T \cdots \Delta \boldsymbol{z}_{mk+m-1}^T \right]^T$

$\overline{\boldsymbol{A}}_L = \boldsymbol{A}_d^m$; $\qquad \overline{\boldsymbol{B}}_L = \left[ \boldsymbol{A}_d^{m-1} \boldsymbol{B}_d \quad \boldsymbol{A}_d^{m-2} \boldsymbol{B}_d \cdots \ \boldsymbol{B}_d \right]$

$$\overline{\boldsymbol{C}}_L = \begin{bmatrix} \boldsymbol{C}_d \\ \boldsymbol{C}_d \boldsymbol{A}_d \\ \boldsymbol{C}_d \boldsymbol{A}_d^2 \\ \vdots \\ \boldsymbol{C}_d \boldsymbol{A}_d^{m-1} \end{bmatrix}; \ \overline{\boldsymbol{D}}_L = \begin{bmatrix} \boldsymbol{D}_d & 0 & \cdots & 0 \\ \boldsymbol{C}_d \boldsymbol{B}_d & \boldsymbol{D}_d & & \vdots \\ \boldsymbol{C}_d \boldsymbol{A}_d \boldsymbol{B}_d & \boldsymbol{C}_d \boldsymbol{B}_d & \boldsymbol{D}_d & \\ \vdots & & \ddots & \ddots & 0 \\ \boldsymbol{C}_d \boldsymbol{A}_d^{m-2} \boldsymbol{B}_d & \boldsymbol{C}_d \boldsymbol{A}_d^{m-3} \boldsymbol{B}_d & \cdots & \boldsymbol{C}_d \boldsymbol{B}_d & \boldsymbol{D}_d \end{bmatrix}$$

where $\overline{\boldsymbol{f}}_K$ and $\Delta \overline{\boldsymbol{z}}_K$ represent the sequences of the cutting force and the vibration over the tooth passing angle $\Theta$, respectively.

Second, the semi-discretized cutting force dynamics over the tooth passing angle $\Theta$ in (12) is lifted as:

$$\overline{\boldsymbol{f}}_K = a_p \left[ \overline{\boldsymbol{r}} - \overline{\boldsymbol{S}} \left( \overline{\boldsymbol{s}}_t + \Delta \overline{\boldsymbol{z}}_K - \Delta \overline{\boldsymbol{z}}_{K-1} \right) \right] \tag{21}$$

where

$$\overline{\boldsymbol{s}}_t = \begin{bmatrix} \boldsymbol{s}_t \\ \boldsymbol{s}_t \\ \vdots \\ \boldsymbol{s}_t \end{bmatrix}; \ \overline{\boldsymbol{r}} = \begin{bmatrix} \boldsymbol{r}_0 \\ \boldsymbol{r}_1 \\ \vdots \\ \boldsymbol{r}_{m-1} \end{bmatrix}; \ \overline{\boldsymbol{S}} = \begin{bmatrix} \boldsymbol{S}_0 & & 0 \\ & \boldsymbol{S}_1 & \\ & & \ddots & \\ 0 & & & \boldsymbol{S}_{m-1} \end{bmatrix}$$

The lifted semi-discretized feedback system is shown in Fig. 2(c). Here, the mechanical system and the cutting process dynamics represented by the discrete-angle shift-invariant system (20) and (21) can be transformed into the standard state equation form for stability and dynamic response analysis:

$$\overline{\boldsymbol{\xi}}_{K+1} = \overline{\boldsymbol{\Phi}} \overline{\boldsymbol{\xi}}_K + \overline{\boldsymbol{\sigma}} \tag{22}$$

where the state vector is $\overline{\boldsymbol{\xi}}_K = [ \overline{\boldsymbol{p}}_K^T \quad \Delta \boldsymbol{z}_{K-1}^T ]^T$;

$$\overline{\boldsymbol{\Phi}} = \begin{bmatrix} \overline{\boldsymbol{A}}_L - a_p \overline{\boldsymbol{B}}_L \left( \boldsymbol{I} + a_p \overline{\boldsymbol{S}} \overline{\boldsymbol{D}}_L \right)^{-1} \overline{\boldsymbol{S}} \overline{\boldsymbol{C}}_L & a_p \overline{\boldsymbol{B}}_L \left( \boldsymbol{I} + a_p \overline{\boldsymbol{S}} \overline{\boldsymbol{D}}_L \right)^{-1} \overline{\boldsymbol{S}} \\ \left( \boldsymbol{I} + a_p \overline{\boldsymbol{D}}_L \overline{\boldsymbol{S}} \right)^{-1} \overline{\boldsymbol{C}}_L & a_p \left( \boldsymbol{I} + a_p \overline{\boldsymbol{D}}_L \overline{\boldsymbol{S}} \right)^{-1} \overline{\boldsymbol{D}}_L \overline{\boldsymbol{S}} \end{bmatrix}$$

$$\overline{\boldsymbol{\sigma}} = \begin{bmatrix} a_p \overline{\boldsymbol{B}}_L \left( \boldsymbol{I} + a_p \overline{\boldsymbol{S}} \overline{\boldsymbol{D}}_L \right)^{-1} \\ a_p \left( \boldsymbol{I} + a_p \overline{\boldsymbol{D}}_L \overline{\boldsymbol{S}} \right)^{-1} \overline{\boldsymbol{D}}_L \end{bmatrix} \left( \overline{\boldsymbol{r}} - \overline{\boldsymbol{S}} \overline{\boldsymbol{s}}_t \right)$$

### D. Stability Analysis

The cutting stability is only induced by the self-excited vibration. The forced term does not affect the stability. Hence, the feedforward part, i.e., $\overline{\boldsymbol{\sigma}}$ in (22), can be neglected in this analysis. Hence, the stability can be determined through the eigenvalues of the monodromy matrix $\overline{\boldsymbol{\Phi}}$:

$$max(abs(eig(\overline{\boldsymbol{\Phi}}))) = \begin{cases} < 1; & \text{stable} \\ = 1; & \text{marginally stable} \\ > 1; & \text{unstable} \end{cases} \tag{23}$$

The monodromy matrix $\overline{\boldsymbol{\Phi}}$ is $r(2n+m)$-dimensional, while that in most of the literature $\boldsymbol{\Phi}_a$ is $r(2n)(m+1)$-dimensional. Since the value of $m$ is in general relatively larger than $n$, the size of $\overline{\boldsymbol{\Phi}}$ is reduced by a factor of approximately $2n$. This size advantage leads to a significant acceleration in eigenvalue computation, as the matrix eigenvalue or inverse computation time scales with the cube of the matrix's dimensions [44].

### E. SLE Prediction

The SLE is one of the indicators to evaluate the surface quality. It is defined as the maximal distance between the desired surface and the machined surface, which can be approximated by the tool/workpiece vibratory motion with the tool geometry. In case of stable cutting, the transient vibration quickly fades away, after the cutter engages the workpiece. Hence, the machined surface can be estimated by the steady-state vibration.

Recall (11), all the existing methods extract the steady-state vibration from $\boldsymbol{\xi}_a^{ss} = (\boldsymbol{I} - \boldsymbol{\Phi}_a)^{-1} \boldsymbol{\sigma}_a$, which demands the monodromy matrix inverse. Meanwhile, the proposed framework can more efficiently compute the steady-state vibration by considering the feedback closed-loop system (20) and (21):

$$\Delta \overline{\boldsymbol{z}}_{ss} = a_p \left[ \overline{\boldsymbol{C}}_L \left( \boldsymbol{I} - \overline{\boldsymbol{A}}_L \right)^{-1} \overline{\boldsymbol{B}}_L + \overline{\boldsymbol{D}}_L \right] \left( \overline{\boldsymbol{r}} - \overline{\boldsymbol{S}} \overline{\boldsymbol{s}}_t \right) \tag{24}$$

where $\Delta \overline{\boldsymbol{z}}_{ss} = [\Delta \boldsymbol{z}_{ss,0}^T \quad \Delta \boldsymbol{z}_{ss,1}^T \quad \cdots \quad \Delta \boldsymbol{z}_{ss,m}^T]^T$. The computation demands the inverse of $\overline{\boldsymbol{A}}_L$ with dimension only $2nr$, which is $m+1$ times smaller than the monodromy matrix in the existing methods. This implies the noticeably improved computation efficiency for the SLE prediction.

The machined surface is a consequence of the cutting edge shearing through the material. It can be determined by the cutter trajectory. Hence, the machined surface, which corresponds to the trajectory of the $j$-th cutting edge during one tooth passing period, involves the interplay of the steady-state tool/workpiece vibration, tool rotation, and feed motion as:

$$\begin{bmatrix} x_{e,j,k} \\ y_{e,j,k} \end{bmatrix} = \Delta \boldsymbol{z}_{ss,k} + \frac{D}{2} \begin{bmatrix} \sin \phi_{j,k} \\ \cos \phi_{j,k} \end{bmatrix} + \frac{k}{m} \boldsymbol{s}_t; k = 0,1,\cdots,m-1 \tag{25}$$

where $\phi_{j,k} = \phi(k\Delta\theta)$. Then, the SLE can be predicted [40] as:

$$SLE = \begin{cases} \dfrac{D}{2} - \max_{j,k}(y_{e,j,k}); & \text{up-milling} \\ -\left\{ \dfrac{D}{2} - \max_{j,k}(y_{e,j,k}) \right\}; & \text{down-milling} \end{cases} \tag{26}$$

It is clearly seen in (24) that the vibration magnitude is linearly proportional to the depth of cut $a_p$ and the feed per tooth $\bar{s}_t$, corresponding to practical experiments. Nonetheless, SLE is nonlinear to the two cutting conditions, due to the nonlinear complex formulas (25) and (26). As the $y$-axial trajectory is periodic, it is sufficient to predict the SLE by one tooth passing angle, i.e., $k = 0, 1, \cdots, m-1$. However, the number of $m$ should be sufficiently large for accurate prediction. The sign of SLE determines whether the cutting is undercut or overcut. Positive SLE indicates undercut, and vice versa. Then, compensation can be made by adjusting the radial depth of cut $a_r$.

The proposed framework integrating the semi-discretization and lifting methods treats the machining system dynamics as a unified discrete-angle linear shift-invariant state-space system model (20) and (21), as depicted in Fig. 2(c). This treatment shows the explicit analytical framework of the system dynamics in the state-space representation, allowing significantly efficient computation of the stability and forced vibration.

**Remark 1**: The main differences between the proposed method and the existing time-domain discretization methods [17-32] lie in three aspects. Firstly, the existing methods discretize the closed-loop model (10) and capture its dynamics over the tooth passing period, directly. The resulting augmented state vector includes all vibratory displacement and velocity states at each discretizing point, rendering the excessive monodromy matrix. In contrast, the proposed method discretizes and lifts the cutting force and mechanical model, separately. As a result, the unnecessary states that do not affected the input-output behavior are dropped from the resulting closed-loop model (22), the proposed method achieves a minimal monodromy matrix, thereby improving computational efficiency in stability analysis. Secondly, most of the semi-discretization methods [17-26] approximate the varying and delayed terms with backward interpolation functions which introduce phase distortion, demanding finer discretization steps for numerical accuracy. Meanwhile, the proposed method implements the zero-phase (or central) interpolation technique. Finally, the separate lifted mechanical model (20) and cutting force model (21) with the feedback structure depicted in Fig. 2(c) enables the steady-state forced vibration to be calculated by the compact model (24), which is more efficient than $\boldsymbol{\xi}_a^{ss} = (\mathbf{I} - \boldsymbol{\Phi}_a)^{-1} \boldsymbol{\sigma}_a$ in the existing methods.

**Remark 2**: The block diagram in Fig. 2(c) developed by the proposed framework provides an explicit understanding of the entire system dynamics and eases the analyses of the system stability and the steady-state response. For the stability analysis, the forced terms are dropped, as they are unaffected in the system stability (23). In case of the steady-state analysis, the regenerative effect is completely inactive, leading to the disappearance of the feedback action. The steady-state response is solely determined by the feedforward dynamics (24).

## IV. Computations and Analyses

In this section, computations and analyses of the SLD and SLE predictions were conducted to demonstrate the efficiency of the proposed methods, including IMP and ZOH, compared to other four well-known methods, namely SDM [19-21], FDM [22, 23, 43], NIM [29, 50], and CCM [31, 32]. For a fair comparison, NIM and CCM were modified to discretize $m$ intervals over the entire tooth passing period, rather than only during the in-cut phase. These computations considered a two-axial ($r = 2$) mechanical system with two participation modes ($n = 2$). Cutting conditions for the SLD and SLE analyses listed in Table 1 were retrieved from [51]. Transfer functions of a mechanical system with $n = 2$ modes of vibrations can be modeled as:

$$G_i(s) = \sum_{j=1}^{2} \frac{\omega_{ni,j}^2 / K_{i,j}}{s^2 + 2\xi_{i,j}\omega_{ni,j}s + \omega_{ni,j}^2}, \quad i = x, y \qquad (27)$$

where $K$ is the structural stiffness, $\xi$ is the damping ratio, and $\omega_n$ is the natural frequency. Equation (27) represents a connected structure between the cutting tool and the workpiece, each stands for a single-mode mass-spring-damper subsystem. In the case of a single mode ($n = 1$), only one 2nd-order transfer function of the primary mode is left in (27). When multiple modes exist on each side, their respective transfer functions must be multiplied together prior to the summation. The transfer function can be converted to the state-space model (9) by any realization technique. The following computations include influences of the number of participation modes $n$, the immersion ratio $a_r/D$ and the discretization steps $m$ to the computation time and prediction accuracy. They were carried out using MATLAB R2023b software on a desktop computer with Intel(R) Core(TM) i7-13700 CPU@2.10 GHz and 16 GB memory.

### A. Rate of Convergence

To demonstrate the accuracy and efficiency of the proposed method, the convergence of $|(\mu(m) - \mu_0)/\mu_0|$ with respect to $m$ was investigated, where $\mu_0$ is the exact eigenvalue and $\mu$ is the estimated eigenvalue as the function of the discretization steps $m$. This comparative study investigated the six different methods, namely 1st SDM [19], 3rd UFDM [23], 2nd INIM [29], and the proposed methods. The exact critical eigenvalue $\mu_0$ was calculated by the IMP with $m = 1000$. This computation was conducted under spindle speed of $\Omega = 4$ krpm, full-immersion milling $a_r/D = 1.0$, and axial depths of cut at $a_p = 0.7, 0.9,$ and $1.1$ mm. The convergence rate of the six methods according to $m$ ranging from 20 to 100 is presented in Fig. 4.

Table 1. Cutting conditions for SLD and SLE analyses

| Cutting Parameters | | Values |
|---|---|---|
| Milling rotational direction | | Down-milling |
| Number of teeth $N$ | | 2 |
| Cutter diameter $D$ (mm) | | 25.0 |
| Feed per tooth $s_x, s_y$ (mm/tooth) | | 0.2/0.0 |
| Tangential/normal cutting coefficients $k_c^t, k_c^n$ (N/mm$^2$) | | 838.7/384.6 |
| Tangential/normal edge coefficients $k_e^t, k_e^n$ (N/mm) | | 19.59/21.18 |
| $X$-axis (1st/2nd modes) | Natural frequency $\omega_{nx}$ (Hz) | 350/540 |
| | Damping coefficients $\xi_x$ | 0.042/0.040 |
| | Structural stiffness $K_x$ (N/$\mu$m) | 38.462/1.681 |
| $Y$-axis (1st/2nd modes) | Natural frequency $\omega_{ny}$ (Hz) | 284/554 |
| | Damping coefficients $\xi_y$ | 0.054/0.190 |
| | Structural stiffness $K_y$ (N/$\mu$m) | 16.129/6.579 |

It is clearly seen in Fig. 4 that $|(\mu(m) - \mu_0)/\mu_0|$ converges to zero in all the cases, as the discretization step $m$ increases. However, there exist noticeable discrepancies in the UFDM and INIM methods, even when $m = 100$. The convergence trajectories between SDM and ZOH are relatively equivalent. The IMP achieves convergence faster than all the time-domain methods, namely SDM, UFDM, INIM, and ZOH. This rapid convergence indicates that the IMP method can reliably estimate stability with smaller $m$ values, indicating its improved efficiency and accuracy in comparison to the other methods. However, the CCM, the spectral method, shows the more rapid convergence over the others since $m = 20$, demonstrating the superior estimation accuracy in the full-immersion cutting.

### B. Stability Lobe Diagram (SLD)

To highlight both the accuracy and efficiency computation of the proposed methods under low immersion milling and higher participation modes, the SLDs were estimated by the same set of methods in Subsection IV-A. The cutting parameters are the same as in Table 1. The reference stability margins were created by the IMP with $m = 300$. The stability charts are calculated over 100×100 sized grids with the spindle speed $\Omega$ ranging from 3 to 23 krpm. The discretization step $m$ is chosen to 20, 30, and 40 with the immersion milling ratio $a_r/D$ of 0.1, 0.5, and 1.0. This computation also studied the mechanical system with one and two participation modes ($n = 1$, 2). For the sake of conciseness, the prediction accuracy and efficiency of the six methods were evaluated by relative error $\varepsilon$ of the stability margins between the candidate methods and the reference, and normalized computation time $t_{norm}$, respectively, depicted in Fig. 5. The relative error $\varepsilon$ and the normalized time $t_{norm}$ are defined as:

$$\varepsilon = \frac{\sum_{i=1}^{d} |\alpha_{ref}(i) - \alpha_{est}(i)|}{\sum_{i=1}^{d} |\alpha_{ref}(i)|} \times 100\%; \quad t_{norm} = \frac{t_{cand}}{t_{base}} \qquad (28)$$

where $\alpha_{ref}(i)$ and $\alpha_{est}(i)$ are the reference and estimated values at the $i^{th}$ data index for spindle speeds, respectively; $d$ is the number of the data; and $t_{cand}$ and $t_{base}$ are the computation times that the candidate and IMP methods spend, respectively. Only the IMP method with $m = 20$ is plotted in Fig. 7. In the figure, each subfigure contains (1) SLD (or stability margin), which is plotted as a solid line isolating the stable cutting zone (color) and the chatter zone (white), and (2) SLE, which is represented by the colormap format within the stability margin. The latter will be discussed in the following subsection.

Fig. 5 shows that the IMP method achieves smaller $\varepsilon$ with shorter computation time in most cases. The SDM offers accurate estimations in all the cases at the expense of a severe computational load. The UFDM achieved the computation faster than the SDM, but estimated inaccurately. The INIM has the better accuracy and speed compared to the SDM and the UFDM, respectively. The CCM outperforms all the methods in only the full-immersion, but is inferior in the interrupted cutting, with heavy computational load. The proposed methods, IMP and ZOH, show comparable results, which are more accurate in most cases. Moreover, they are more efficient than all the other methods. In cases of the one mode, the proposed methods are faster than the others up to 10 times, while, in the two modes, it is much faster up to 40 times. This signifies the uniform great accuracy in low and full immersion cuttings and the noticeable improvement in efficiency of the proposed methods.

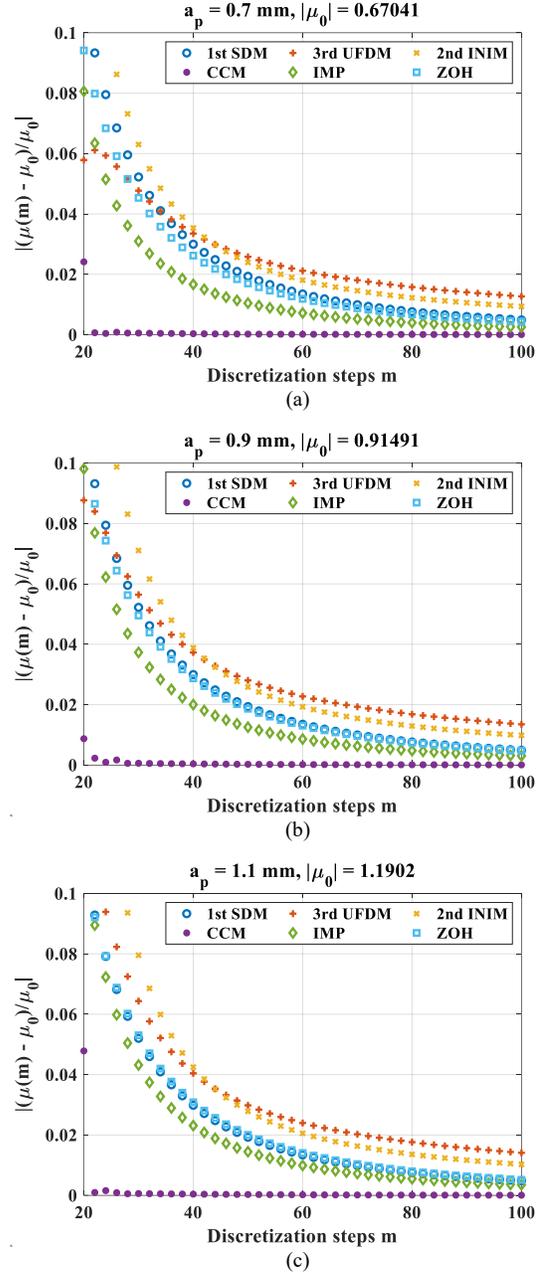

Fig. 4. Convergence rate of the critical eigenvalues

Fig. 7 illustrates the SLDs estimated by the proposed method with only $m = 20$. The system with two modes ($n = 2$) shows much smaller critical depth of cut $a_p$ than those with one dominant mode ($n = 1$), indicating that the non-dominant modes may contribute to unstable cutting at depths of cut, where the analysis of the one-mode system determines stable cutting. Hence, in some particular cases, cutting stability analyses with only one vibration mode are not as accurate as those with multiple modes included. However, numerous existing methods require abundant time to complete the analyses, but the proposed methods do not.

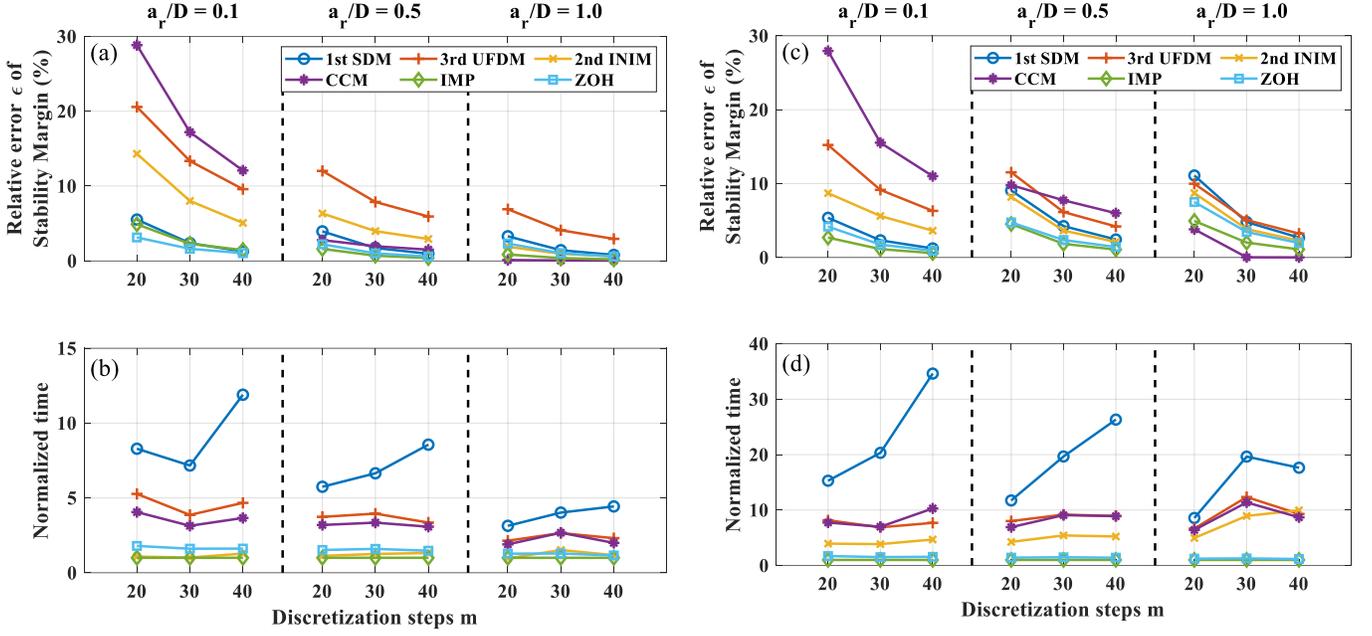

Fig. 5. Relative error of stability margin and normalized computation time: (a-b) one participation mode ($n = 1$), and (c-d) two participation modes ($n = 2$)

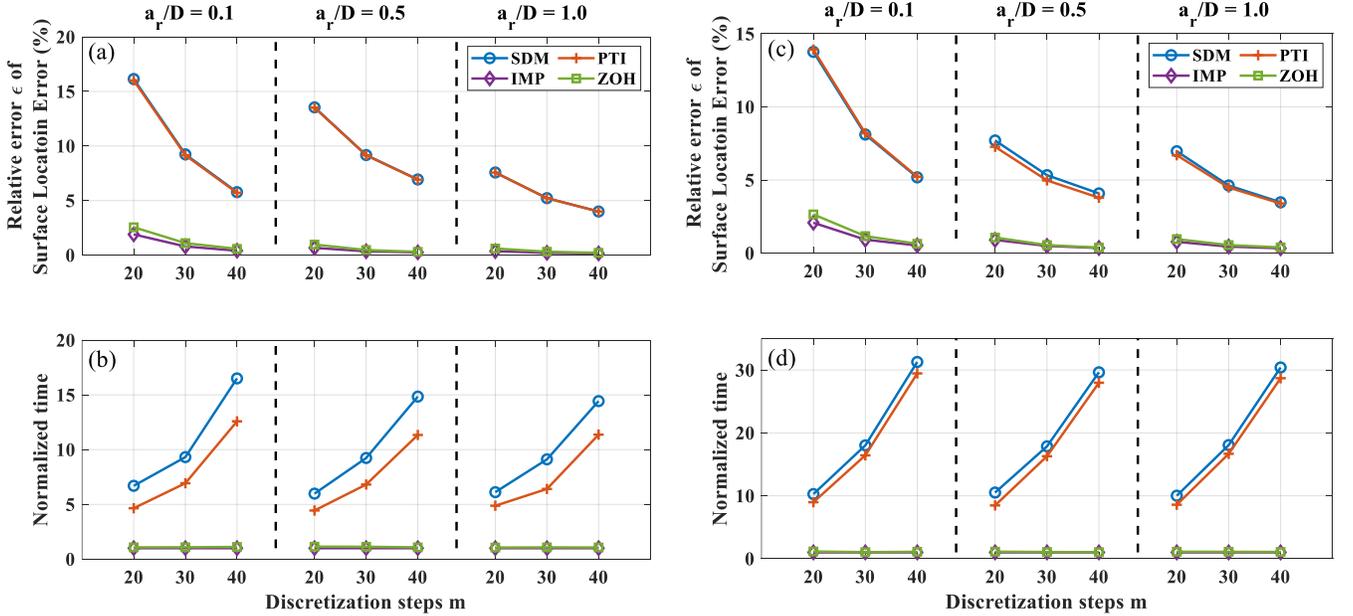

Fig. 6. Relative error of SLE and normalized computation time: (a-b) one participation mode ($n = 1$), and (c-d) two participation modes ($n = 2$)

### C. Surface Location Error (SLE)

In addition to the stability estimation, the proposed methods are capable of predicting SLE. Two SLE estimation approaches, i.e., 2[nd] SDM [20] and PTI [43], were used to highlight the performance of the proposed method in predicting SLE. The PTI is an extended version of the FDM approach, particularly for the SLE analysis. The cutting parameters are the same as for Table 1. The dept of cut $a_p$ is set to 0.5 mm, which provides stable cutting for all the cases. The SLE prediction resulted from the IMP with $m = 1000$ was taken as the reference. The SLE was investigated over the spindle speed ranging between 3 and 23 krpm sliced into 200 grids. Analogous to the SLD case, the discretization step $m$ is chosen to 20, 30, and 40 with the immersion milling ratio $a_r/D$ of 0.1, 0.5, and 1.0. This study case also considered the mechanical system with one and two participation modes ($n = 1$, 2). To evaluate the accuracy and efficiency, the $\varepsilon$ of SLE and $t_{norm}$ are illustrated in Fig. 6.

Fig. 6 presents that the agreements tend to be better in the full immersion than the low immersion milling. The estimation performances of the SDM and PTI methods are comparable in terms of accuracy and efficiency. Analogously, the performances of the IMP and ZOH are also equivalent. Compared to the SDM and PTI methods at the same $m$, the proposed methods have significantly smaller $\varepsilon$ and their computation times are also shorter by 6 up to 17 times in the one-mode case, and by 10 up to 32 times in the two-mode case. Among the proposed methods, the IMP outperforms the ZOH. On the other hand, SLEs under the stability margins estimated by the IMP with $m = 20$ are presented as colormaps in Fig. 7. Non-solid contour lines indicate different levels of SLEs. In other words, the SLE remains constant along each contour line. Spindle speeds where contour lines meet show transitions from undercut (positive SLE) to overcut (negative SLE), or vice versa. As evident from (24)-(26), increasing the depth of cut $a_p$ or the feed per tooth $s_t$ amplifies the tool/workpiece vibration, implying a larger SLE. However, given $a_p$ and $s_t$, the SLE varies along spindle speed. The relatively large SLEs are observed near the subharmonics of the natural frequency $f_n$. The spindle speeds where the large SLEs occur can be estimated by:

$$\Omega_{SLE} = \frac{60 f_n}{kN}; \quad k = 1, 2, \ldots \quad (29)$$

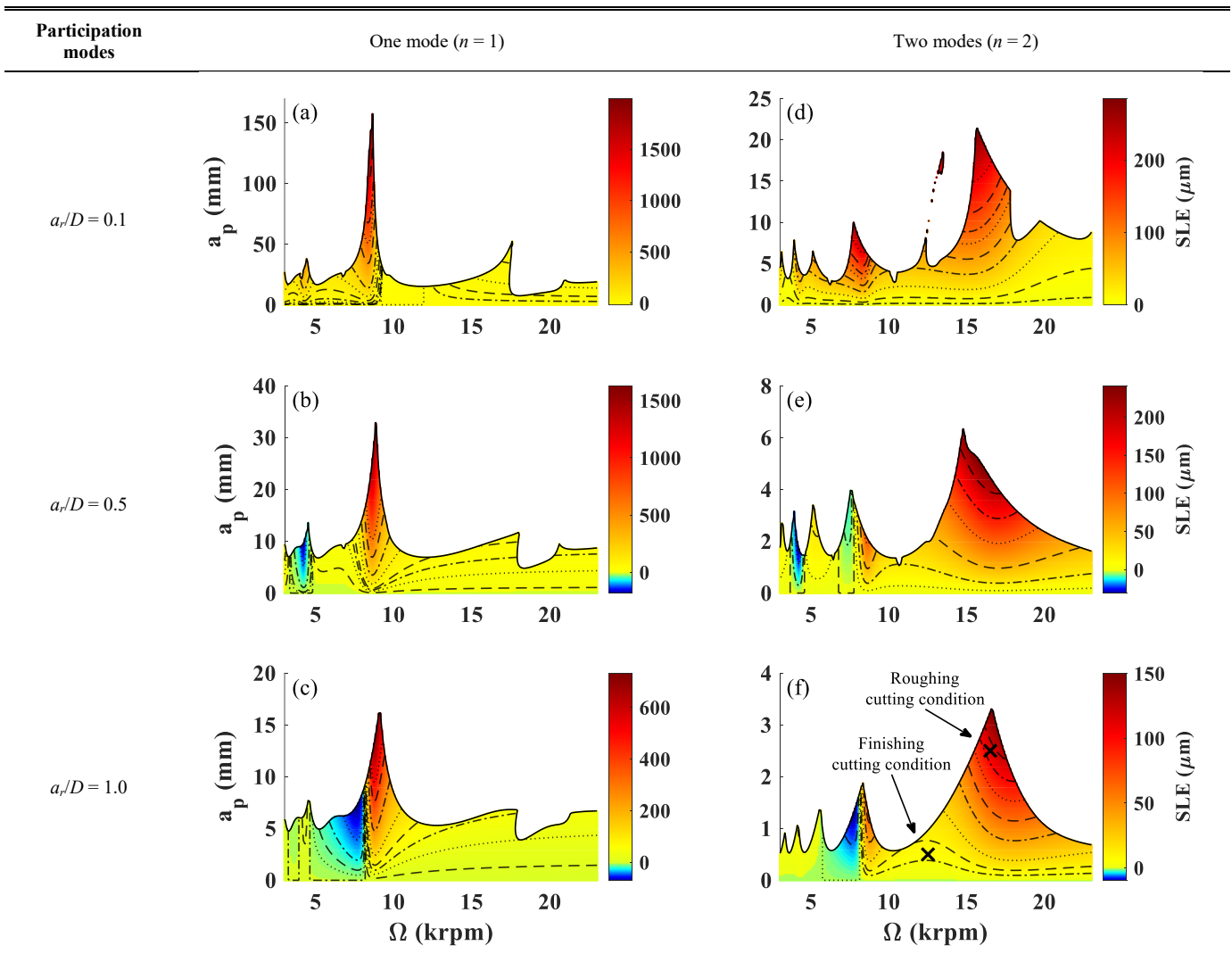

Fig. 7. SLD-SLE contour diagram with the proposed method for the mechanical system with (a-c) one mode and (d-f) two modes.

The large SLEs occur due to higher subharmonics under the lower immersion cutting. Hence, the choice of the cutting conditions, i.e., $a_p$ and $\Omega$, should consider not only the SLD, but also the SLE colormap, in order to achieve stable cutting along with good surface quality. It will be further discussed in the following subsection.

To comprehend the influence of the tool/workpiece vibration to the SLE, the vibration over one tooth passing angle with respect to three different spindle speeds $\Omega$, i.e., 4.2, 7.4, and 8.6 krpm, and three discretization steps $m$, i.e., 20, 30, and 40, are plotted in Fig. 8. The computation with the immersion ratio of $a_r/D = 0.5$ was studied under the one-mode case. It is found that the minimal SLE is contributed by a tiny vibrational magnitude and a 90-degree out-of-phase alignment with the cutting edge's engagement. It is intuitively understood that a smaller vibrational magnitude implies a tinier SLE when comparing between the blue (circle) and yellow (diamond) lines. However, despite the blue and red (square) lines having indifferent magnitudes, the red line leads to a much smaller SLE. This is due to the fact that the cutter engagement follows the cosine function of the cutter angular position (second line in (25)), requiring zero vibration at the deepest engagement angles (i.e., $\theta = 2i\pi/N$, $i = 0$, 1, 2, ...) for the minimum SLE. On the other hand, the discretization step $m$ determines the number of samples on the vibration profiles, as observed in Fig. 8. Hence, increasing *m* improves the accuracy of the SLE prediction.

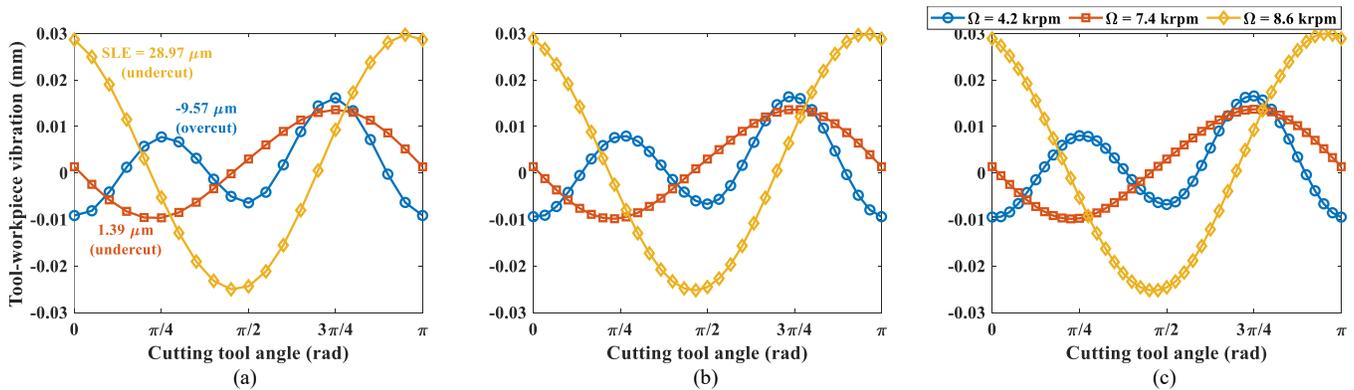

Fig. 8. Prediction of the *y*-axial tool/workpiece vibration over one tooth passing angle $\Delta y_{ss}$ with the discretization steps *m* of (a) 20, (b) 30, and (c) 40

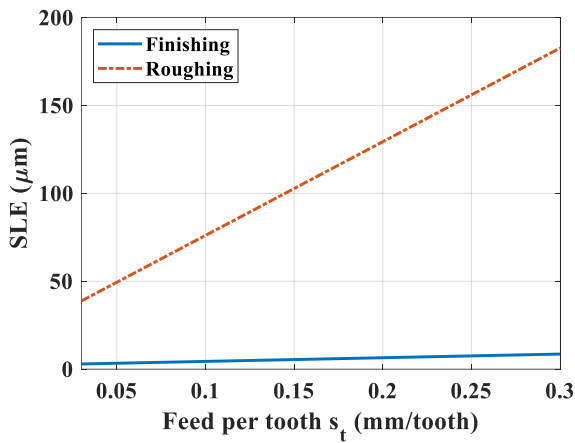

Fig. 9. SLEs of the roughing and finishing cutting conditions for different feed per tooth $s_t$

### D. SLD and SLE Analyses and Discussions

As depicted in the previous subsections, the proposed methods, especially the IMP method, show the outstanding prediction, especially efficiency, over the existing methods. For the SLD analysis, the main reason behind this superior efficiency lies on the reduced size of the monodromy matrix. Recall that the monodromy matrix in all the SDM, UFDM, INIM, and CCM is $r(2n)(m+1)$-dimensional. On the other hand, that of the proposed minimal realization method is only $r(2n+m)$-dimensional. Under the specified cutting conditions, the size of the monodromy matrix in the proposed method is approximately 4 times smaller. Computation of the eigenvalues of the monodromy matrix is dominant in the stability analysis. The runtime of the eigenvalue computation grows as the cubic power of the matrix size [44]. For this reason, increasing the number of modes *n* or discretization steps *m* results in significantly greater differences in computational loads.

For the SLE analysis, all the proposed methods are also considerably more accurate and efficient than the existing methods, due to two factors. Firstly, the proposed methods are based on the zero-phase conversion and interpolation, avoiding the estimation inaccuracy due to the phase delay. Secondly, the existing methods calculate the steady-state vibration $\Delta \bar{z}_{ss}$ by means of the monodromy matrix, which is large in size and may require iterative matrix multiplications. It is much more time-consuming than the proposed methods, in which the calculation involves only the lifted system matrices (24), which are more compact than the monodromy matrix. As a result, the proposed methods can efficiently achieve the accurate SLE prediction by a merely small value of *m*, whereas the other methods require plenty of time and a larger of *m*.

Fig. 7 presents the so-called SLD-SLE contour chart, where SLEs are visualized in the colormap format inside the stability margin, while the (white) area outside the margin indicates the unstable cutting. This chart is beneficial in prudently selecting proper cutting conditions for the purposes of avoiding chatter in roughing and minimizing SLE in finishing cut. For instance, in the case study shown in Fig. 7(f), a depth of cut of $a_p = 2.5$ mm and spindle speed of $\Omega = 16.5$ krpm can be selected to maximize the material removal rate in roughing cut. For finishing cut, the depth of cut may be limited to $a_p = 0.5$ mm to ensure the stable cutting; therefore, the spindle speed can be selected to $\Omega = 12.5$ krpm. By the two cutting conditions, the SLEs corresponding to feed per tooth $s_t$ ranging from 0.03 to 0.3 mm/tooth are plotted in Fig. 9. The SLE is observed to be linearly proportional to the feed per tooth within this range. Furthermore, the results also emphasize that the finishing cutting condition offers significantly smaller SLEs compared to the roughing one. Although zero SLE can be achieved at the transition locations, the feasible spindle speed range is too sensitive and narrow. Furthermore, its neighborhood is characterized by fluctuations of SLEs. Thus, choosing the spindle speed at the transition zone is too risky for consistent and reliable cutting.

### V. CONCLUSIONS

This paper presents a novel framework to model and analyze complex machining dynamic systems for a wide range of machining processes. The framework integrates zero-phase SDM, state space minimal realization of the feedback system, and the lifting method on the periodic varying dynamics, establishing the explicit dynamic system in the discrete-angle linear shift-invariant state-space form that allows for efficient computation for the stability analysis and SLE prediction. The two-dimensional milling operation with cutter dynamics is taken as the study case examples. With one and two vibration modes on each motion axis, the results demonstrate the significant improvement of the proposed method over the existing methods in the number of discretization steps and the computation load, while rendering similar accuracy. The SLD-SLE contour charts in the example provide visualization for selecting cutting parameters for multi-pass cutting in avoiding chatter in roughing cut and minimizing SLE in finishing cut.

Finally, the proposed dynamic system framework along with the analyses lends itself to addressing complex machining situations. For example, complex workpiece geometry and dynamics may be modeled using FEM and included in the mechanical model, establishing a more general machining dynamic system. The dynamics may vary with respect to the cutter-workpiece interface locations, leading to different stability and vibration profiles throughout the workpiece surface. Furthermore, the semi-discretization intervals are not necessarily uniform as considered in this paper. The framework can readily analyze the situation of variable spindle speed or uneven tooth spacing.

## APPENDIX A

This section shows the derivations of the zero-phase continuous-to-discrete conversion for IMP (18) and ZOH (19). The discrete-time general solution of (9) is written as:

$$\boldsymbol{q}_{k+1} = e^{\boldsymbol{A}_\omega \Delta\theta} \boldsymbol{q}_k + \int_0^{\Delta\theta} e^{\boldsymbol{A}_\omega \gamma} \boldsymbol{B}_\omega \boldsymbol{f}((k+1)\Delta\theta - \gamma) d\gamma \quad (30)$$

$$\boldsymbol{y}_k = \boldsymbol{C}_\omega \boldsymbol{q}_k$$

### A. Impulse Invariance Hold (IMP)

For IMP, plugging (14) into (30), one gets

$$\boldsymbol{q}_{k+1} = e^{\boldsymbol{A}_\omega \Delta\theta} \boldsymbol{q}_k + \int_0^{\Delta\theta} e^{\boldsymbol{A}_\omega \gamma} \boldsymbol{B}_\omega \boldsymbol{f}_k \delta(\Delta\theta - \gamma)\Delta\theta d\gamma \quad (31)$$

$$= e^{\boldsymbol{A}_\omega \Delta\theta} \boldsymbol{q}_k + e^{\boldsymbol{A}_\omega \Delta\theta} \boldsymbol{B}_\omega \Delta\theta \boldsymbol{f}_k$$

Redefining the state by applying (17), (33) is rewritten as:

$$\boldsymbol{p}_{k+1} = \underbrace{e^{\boldsymbol{A}_\omega \Delta\theta}}_{\boldsymbol{A}_d} \boldsymbol{p}_k + \underbrace{e^{\boldsymbol{A}_\omega \Delta\theta} \boldsymbol{B}_\omega \Delta\theta}_{\boldsymbol{B}_d} \boldsymbol{f}_k$$

$$\boldsymbol{y}_k = \underbrace{\boldsymbol{C}_\omega}_{\boldsymbol{C}_d} \boldsymbol{p}_k \quad (32)$$

### B. Zero-phase Zero-order Hold (ZOH)

For ZOH, plugging (15) into (30), one gets

$$\boldsymbol{q}_{k+1} = e^{\boldsymbol{A}_\omega \Delta\theta} \boldsymbol{q}_k + \int_0^{\frac{\Delta\theta}{2}} e^{\boldsymbol{A}_\omega \gamma} \boldsymbol{B}_\omega \boldsymbol{f}_{k+1} d\gamma + \int_0^{\frac{\Delta\theta}{2}} e^{\boldsymbol{A}_\omega \gamma} \boldsymbol{B}_\omega \boldsymbol{f}_k d\gamma$$

$$= e^{\boldsymbol{A}_\omega \Delta\theta} \boldsymbol{q}_k + \underbrace{\left(e^{\boldsymbol{A}_\omega \frac{\Delta\theta}{2}} - \boldsymbol{I}\right) \boldsymbol{A}_\omega^{-1} \boldsymbol{B}_\omega}_{\boldsymbol{E}_d} \boldsymbol{f}_{k+1} + e^{\boldsymbol{A}_\omega \frac{\Delta\theta}{2}} \left(e^{\boldsymbol{A}_\omega \frac{\Delta\theta}{2}} - \boldsymbol{I}\right) \boldsymbol{A}_\omega^{-1} \boldsymbol{B}_\omega \boldsymbol{f}_k \quad (33)$$

Redefining the state by applying (17), (33) is rewritten as:

$$\boldsymbol{p}_{k+1} = e^{\boldsymbol{A}_\omega \Delta\theta} \left(\boldsymbol{p}_k + \boldsymbol{E}_d \boldsymbol{f}_k\right) + e^{\boldsymbol{A}_\omega \frac{\Delta\theta}{2}} \boldsymbol{E}_d \boldsymbol{f}_k$$

$$= \underbrace{e^{\boldsymbol{A}_\omega \Delta\theta}}_{\boldsymbol{A}_d} \boldsymbol{p}_k + \underbrace{e^{\boldsymbol{A}_\omega \frac{\Delta\theta}{2}} \left(e^{\boldsymbol{A}_\omega \frac{\Delta\theta}{2}} - \boldsymbol{I}\right) \boldsymbol{A}_\omega^{-1} \boldsymbol{B}_\omega}_{\boldsymbol{B}_d} \boldsymbol{f}_k \quad (34)$$

$$\boldsymbol{y}_k = \underbrace{\boldsymbol{C}_\omega}_{\boldsymbol{C}_d} \boldsymbol{p}_k + \underbrace{\boldsymbol{C}_\omega \boldsymbol{E}_d}_{\boldsymbol{D}_d} \boldsymbol{f}_k$$